# General lossless polarization and phase transformation using bilayer metasurfaces

*Babak Mirzapourbeinekalaye, Andrew McClung, and Amir Arbabi\**

*Department of Electrical and Computer Engineering, University of Massachusetts Amherst, 151 Holdsworth Way, Amherst, MA 01003, USA*

*E-mail: arbabi@umass.edu*

Metasurface optical elements enable wavefront control and polarization manipulation with subwavelength resolution. Metasurfaces made of linearly birefringent meta-atoms such as rectangular nano-posts are commonly used to control phase and polarization, but a single layer of these meta-atoms cannot implement transformations with elliptically polarized or circularly polarized eigenstates (i.e., chiral transformations). Here, it is shown that two cascaded metasurface layers comprising linearly birefringent meta-atoms provide sufficiently many degrees of freedom to implement arbitrary unitary phase and polarization transformations. A systematic design method for such metastructures is described and used to design a bifocal metalens that cannot be implemented by a single layer of birefringent meta-atoms. The presented design and implementation techniques introduce a systematic approach for realizing the most general form of lossless polarization and phase transformations using metasurfaces with high efficiency.

# 1. Introduction

Dielectric metasurfaces, prized for their low weight, planar form factor, and potential for low-cost manufacture, are strong candidates for the next generation of optical systems.[1] Beyond compactness, another feature distinguishing metasurface-based components from conventional counterparts is their ability to multiplex different optical functions. Multi-functional components implement different phase profiles for incident fields with different properties (e.g., different wavelengths,[2,3] incidence angles,[4] or polarizations[5]). These components require more degrees of freedom than those with a single function.[6–8]

Previous demonstrations of polarization-multiplexed metasurfaces have acquired these degrees of freedom using meta-atoms which, like waveplates made of quartz, have linearly polarized eigenstates.[5,9–14] Here we refer to such meta-atoms as linearly birefringent (LB) meta-atoms. LB meta-atoms have simple shapes, and metasurfaces formed of them can impose arbitrary wavefronts on incident fields with orthogonal polarizations.[5] However, these metasurfaces cannot simultaneously control the polarization of the transmitted fields, which are constrained to have the same polarization ellipse as their corresponding incident fields but the opposite handedness.[5] Transformations with linearly polarized eigenstates constitute only a subset of the possible unitary transformations.[5] Notably, chiral transformations lie outside this subset and are typically implemented by meta-atoms with more complex shapes.[15–19]

Another way to create multi-functional components is to use cascaded layers of meta-atoms.[3,8,20–24] Here, we show that all lossless phase and polarization transformations can be implemented by two layers of LB meta-atoms. A device with no absorption or reflection, which we refer to as lossless, is described by a unitary transformation. Metasurfaces composed of LB meta-atoms can be highly efficient, straightforwardly designed, and readily fabricated;[5] thus, the implementation of general polarization transformations using a cascade of such metasurfaces is desirable. As a demonstration, we present a design for a polarization-multiplexed metalens that cannot be implemented by a single layer of LB meta-atoms. Using supercells of cascaded rectangular meta-atoms, the metalens transforms right- and left-circularly polarized incident fields into transmitted fields with elliptical polarizations and independent phase profiles, focusing them to two different points.

## 2. Bilayer LB Decomposition

The polarization and phase of polarized monochromatic light and the action of linear elements that transform them are commonly characterized using the Jones matrix formalism. Here we represent the polarization and phase of a plane wave propagating along the $z$ direction by its normalized electric field $\boldsymbol{e} = \sqrt{n/(2\eta_0)}\boldsymbol{E}$, where $n$ is the refractive index of the material that light is propagating in, $\eta_0$ is the free space impedance, and $\boldsymbol{E} = \begin{bmatrix} E_x \\ E_y \end{bmatrix}$ is the phasor of the electric field vector of light assuming a time dependence of $e^{-i\omega t}$, where $\omega$ is the angular frequency of light. Using this normalization, the power per unit area of light is equal to $|\boldsymbol{e}|^2$. In the Jones matrix formalism, the transformation of any linear optical element is described by $\boldsymbol{e}_{\text{out}} = \mathbf{T}\boldsymbol{e}_{\text{in}}$, where $\boldsymbol{e}_{\text{in}}$ and $\boldsymbol{e}_{\text{out}}$ are the normalized electric field vectors of the incident and output light, respectively, and $\mathbf{T}$ is the 2×2 Jones matrix. For simplicity, we assume that the power density of the incident light is normalized to one, thus $|\boldsymbol{e}_{\text{in}}| = 1$. The Jones matrix of a lossless element is unitary. Therefore, its eigenvectors $\boldsymbol{e}_1$ and $\boldsymbol{e}_2$, which we refer to as polarization eigenstates or eigenstates for brevity, are orthogonal, and its eigenvalues have unity modulus and can be written as $e^{i\phi_1}$ and $e^{i\phi_2}$, where $\phi_1$ and $\phi_2$ are real. LB optical elements (e.g., waveplates made of quartz crystal) constitute a subset of the unitary transformations, namely, those whose eigenstates are linearly polarized, and transformations with general polarization eigenstates (i.e., elliptical or circular birefringence) cannot be implemented by a single LB element. Equivalently stated, the Jones matrix of an LB element is symmetric, but that of a general transformation is not.

We can visualize the action of polarization transforming elements using the Poincaré sphere.[25] On the Poincaré sphere, orthogonal states of polarization form antipodal pairs, i.e., a straight line connecting them passes through the sphere's origin. We refer to the line connecting a unitary transformation's eigenstates as the eigenaxis. Polarization transformation by an LB optical element is visualized in **Figure 1a**: its eigenaxis that connects two linearly polarized eigenstates $\boldsymbol{e}_1$ to $\boldsymbol{e}_2$ lie along the Poincaré sphere's equator[26] with an azimuth angle $2\theta$, where $\theta$ is the angle between the optical axis of the LB element and the $x$ axis (Figure 1a, top). An incident field with arbitrary polarization can be represented as a linear combination of the LB element's eigenstates (i.e., $\boldsymbol{e}_{\text{in}} = c_1\boldsymbol{e}_1 + c_2\boldsymbol{e}_2$, where $c_1, c_2 \in \mathbb{C}$). As the light passes through the element, the phase of each component is retarded independently, producing light with a new polarization state

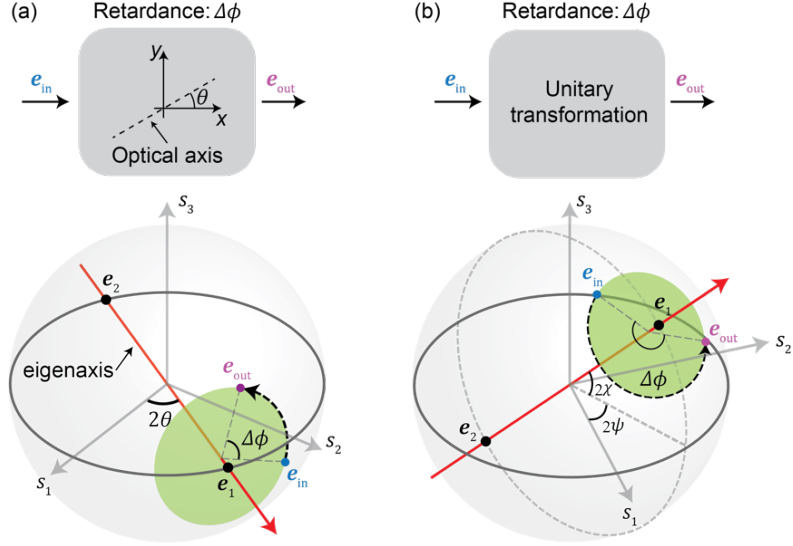

**Figure 1.** a) Relation between the input and output polarization states of an LB transformation. The eigenaxis connects linear eigenstates $e_1$ and $e_2$ which lie on the equator. b) Relation between the input and output polarization states of a unitary transformation with elliptical eigenstates $e_1$ and $e_2$. The transformations in (a) and (b) rotate an input state $e_{in}$ around the eigenaxis by an angle $\Delta\phi = \phi_2 - \phi_1$ to an output state $e_{out}$.

$$e_{out} = e^{i\phi_1}c_1 e_1 + e^{i\phi_2}c_2 e_2 = e^{i\bar{\phi}}(e^{-\frac{i\Delta\phi}{2}}c_1 e_1 + e^{\frac{i\Delta\phi}{2}}c_2 e_2), \quad \text{where} \quad \Delta\phi = \phi_2 - \phi_1 \text{ is the}$$

retardance, and we refer to $\bar{\phi} = (\phi_2 + \phi_1)/2$ as the average phase.

This transformation is represented as a right-handed rotation around the eigenaxis from $e_2$ to $e_1$ by an angle equal to $\Delta\phi$ on the Poincaré sphere (Figure 1a).[26] A polarization and phase transformation by an LB element is fully specified by three parameters: $\theta$, $\Delta\phi$, and $\bar{\phi}$. The average phase only causes a phase shift of the output state and is not represented on the Poincaré sphere.

A general lossless element has elliptical polarization eigenstates that do not necessarily lie on the Poincaré sphere's equator. We assume that $e_1$ has azimuth and elevation angles of $2\psi$ and $2\chi$, respectively (Figure 1b). The action of such an element is also shown in Figure 1b and is the rotation of the incident polarization by the angle $\Delta\phi$ around the eigenaxis.[26] Though this transformation cannot be implemented by a single LB element, it can be implemented by a cascade of two, as we show in the Supporting Information.

As shown in the Supporting Information, the average phase of the cascade of two elements is the sum of the average phases of the individual elements. Thus, when implementing a general lossless element by cascading two LB elements, the average phase of one of the LB elements can be freely selected. Furthermore, there is an additional degree of freedom in the polarization transformation. In the following, we will use the Poincaré sphere representation to illustrate this second degree of freedom and provide an intuitive graphical approach that reveals some of the features of this implementation.

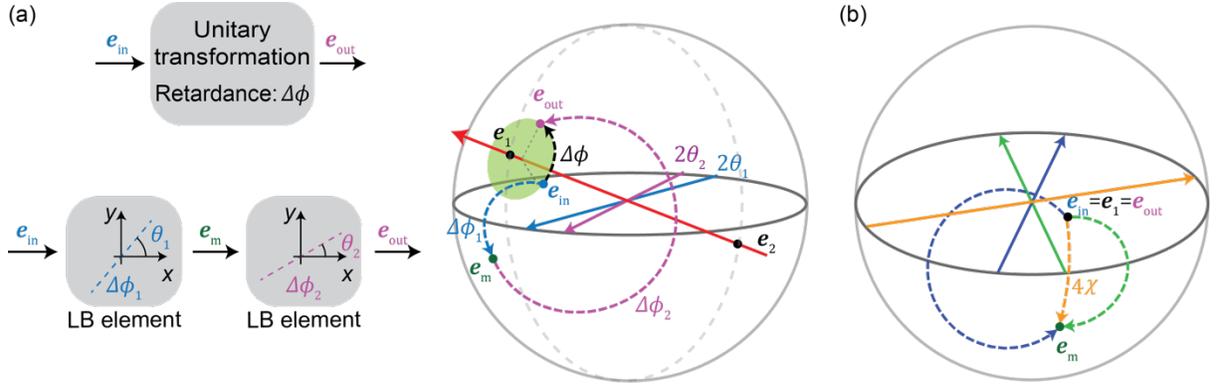

**Figure 2.** a) (Left) Illustration of a unitary transformation and its LB decomposition. (Right) Geometrical illustration of a unitary transformation that rotates an input polarization $e_{\text{in}}$ about an elliptical eigenaxis by an angle $\Delta\phi$ to an output polarization $e_{\text{out}}$. The same operation can be decomposed into two LB operations (blue and purple) that traverse through an intermediate state $e_{\text{m}}$. b) Illustration of three different solutions for the first step of the LB decomposition of a unitary transformation illuminated by an eigenpolarization. The intermediate state is the mirror state of the input polarization with respect to the equatorial plane.

The decomposition of a transformation with elliptical eigenstates into two LB transformations is illustrated in **Figure 2a**: The first LB element rotates the incident polarization around an equatorial eigenaxis with azimuth $2\theta_1$ by an angle $\Delta\phi_1$ (the first element's retardance), bringing an incident field $e_{\text{in}}$ to an intermediate state $e_{\text{m}}$. The second LB element then rotates the polarization state around a second equatorial eigenaxis with azimuth $2\theta_2$ by an angle $\Delta\phi_2$, arriving at an output state $e_{\text{out}}$. The composition of these two rotations results in an effective rotation around an elliptical (non-equatorial) eigenaxis.

We can illustrate the degree of freedom available for the LB decomposition and some universal features of these decompositions by considering a bilayer element illuminated by one of its elliptical eigenstates $e_1$ (Figure 2b). Because $e_1$ is an eigenstate, the output light must have the same polarization as the incident one. Infinitely many pairs of LB elements satisfy this requirement; a few solutions for the first element are shown in Figure 2b. However, in all solutions, the intermediate state $e_{\text{m}}$ is fixed and is the mirror image of $e_1$ with respect to the equatorial plane. This is because the polarization rotation planes for both LB elements are normal to the equatorial plane; thus, $e_1$ and its mirror image with respect to the equatorial plane are the only two states that are on both rotation planes. As a result, the first LB element must transfer $e_1$ to its mirror image, but its rotation plane and corresponding eigenaxis may be freely selected. Selection of the eigenaxis (i.e., $\theta_1$) determines the rotation angle $\Delta\phi_1$ required for transferring $\boldsymbol{e}_1$ to $\boldsymbol{e}_m$. A similar operation is performed by the second LB element which transfers $\boldsymbol{e}_m$ to $\boldsymbol{e}_1$; thus $\theta_2$ determines $\Delta\phi_2$.

Bilayer elements with the same eigenstate $\boldsymbol{e}_1$ and $\theta_1$ but different values of $\Delta\phi$ should have different values of $\theta_2$, otherwise, they will correspond to the same transformation on the Poincaré sphere. Therefore, $\theta_1$ and $\theta_2$ are related by the bilayer element's retardance and only one of the four parameters $\theta_1, \theta_2, \Delta\phi_1$ and $\Delta\phi_2$ can be freely selected. In addition, as it can be seen from the graphical illustration shown in Figure 2b, the retardance values of both LB elements is a value between $4\chi$ and $2\pi - 4\chi$, where $2\chi$ is the elevation angle of the eigenstate $\boldsymbol{e}_1$ on the Poincaré sphere (see Figure 1b) and must be within this range when selected as the degree of freedom. For example, bilayer chiral elements, whose eigenstates are circularly polarized and are located at the sphere poles, can only be realized using two LB elements functioning as half-wave plates because $\chi = \frac{\pi}{2}$ for $\boldsymbol{e}_1$ which leads to $\Delta\phi_1 = \Delta\phi_2 = \pi$.

To summarize, there are two residual degrees of freedom in LB decomposition: one in the average phase of the two LB elements and another in the polarization transformation (i.e., one of the four parameters $\theta_1, \theta_2, \Delta\phi_1$ and $\Delta\phi_2$ can be freely selected). These degrees of freedom can be selected to meet additional design criteria, e.g., to optimize the efficiency of a given design, as we show through design examples in section 4.

## 3. Bilayer Design Method

We now illustrate how the LB decomposition technique can be used to design bilayer metasurfaces that implement the most general form of spatially varying polarization and phase transformation. A schematic illustration of a metasurface composed of an array of cascaded LB elements, each implementing an arbitrary unitary transformation, is shown in **Figure 3a**. To design a spatially varying metasurface, we decompose the unitary Jones matrix of each bilayer metasurface unit cell into two unitary and symmetric Jones matrices and determine their retardances and eigenaxis angles by using Equations A7, A8 and A9, in the Supporting Information and by selecting the two degrees of freedom discussed in section 2. Two LB elements with the corresponding retardances and eigenaxis angles are then implemented using two meta-atoms and are stacked to form the bilayer unit cell. In addition to the local periodic assumption,[5] this partitioning approach ignores potential interlayer couplings among the meta-atoms, and we will study their effects on the device

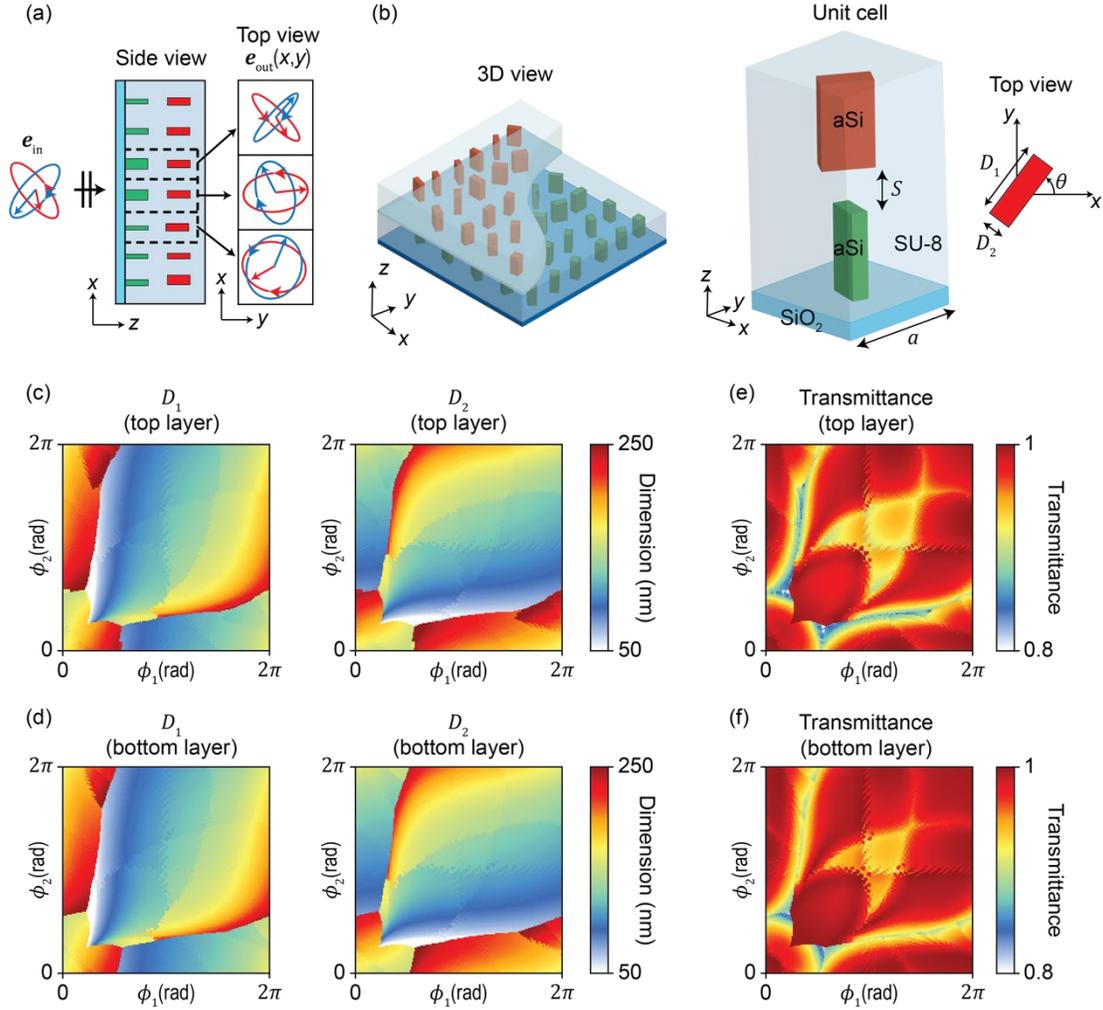

**Figure 3.** a) Schematic illustration of a bilayer metasurface implementing a spatially-varying unitary operation. Blue and red ellipses show the orthogonal polarizations and arrows show their phases at the output of each cell. b) 3D view of a bilayer metasurface and a unit cell. c) Design curves relating post dimensions ($D_1$ and $D_2$) to different phase shifts ($\phi_1$ and $\phi_2$) in the top layer. d) Design curves for the bottom layer. e) The transmittance of unpolarized light associated with (c), and f), with (d).

efficiency using design examples later in this section and in section 4.

Figure 3b shows 3D views of an example implementation of the bilayer unit cells using high refractive index nano-posts with rectangular cross-sections. It has been previously shown that a single layer of such nano-posts with properly selected dimensions and rotation angle can efficiently realize LB elements.[5] In the example implementation, we use an operation wavelength of 850 nm and assume each unit cell consists of two vertically stacked 600-nm-tall amorphous silicon (aSi, $n = 3.82$) nano-posts. The nano-posts are located on a square lattice with 480 nm lattice constant. The bottom layer rests on a fused silica ($n = 1.45$) substrate, and the top layer lies at a distance $S$ from the bottom layer (Figure 3b). The infinite half-space above the fused silica substrate and around the nano-posts is assumed to be filled with SU-8 polymer ($n = 1.56$).

The design maps for each layer are shown in Figure 3c and 3d. These maps are obtained using the method detailed in Ref.[5] and provide optimal values of nano-posts cross-sectional dimensions ($D_1$ and $D_2$) for achieving any desired phase shifts ($\phi_1$ and $\phi_2$) for light polarized along $D_1$ and $D_2$, respectively. Using a workstation (Intel Xeon E5-2680 CPU) and running the unit cell simulations in parallel with 10 cores, the complete design maps can be generated in less than 70 minutes. The transmittances of the layers for unpolarized light are shown in Figure 3e and 3f and are higher than 80% at all points in the design domain.

As mentioned above, the partitioning approach assumes the nano-posts within each bilayer unit cell act independently and neglects interlayer couplings due to multiple reflections and near-field couplings between the layers. To study the effect of interlayer couplings, we consider a periodic array of bilayer supercells in which the interlayer interactions are incorporated by simulating both constituents of the unit cell together **(see Figure 4a)**. While these simulations more accurately capture the physics of our proposed system than the partitioned approach described above, the large number of simulations needed to cover the entire design space precludes the use of the supercell approach to design bilayer metasurfaces. To determine the validity and estimate the error of our partitioned bilayer approximation, we randomly selected a large number (20,000) of unitary Jones matrices $\mathbf{T}^U$ and implemented them using the maps shown in Figure 3 by randomly selecting values for the rotation angle $\theta_1$ and average phase $\bar{\phi}_1$ of the first LB element as degrees of freedom. We then randomly selected incident fields $\boldsymbol{e}_{\text{in}}$ and obtained the accurate output field $\tilde{\boldsymbol{e}}_{\text{out}}$ using full-wave simulations[27] of the bilayer supercells. Because of the smaller than unity transmittance of the nano-posts, the approximate realization of $\phi_1$ and $\phi_2$, and errors introduced by neglecting interlayer couplings, the supercell response $\tilde{\boldsymbol{e}}_{\text{out}}$ differs from the desired output field $\boldsymbol{e}_{\text{out}} = \mathbf{T}^U \boldsymbol{e}_{\text{in}}$.

To quantify the effect of these errors on a metasurface device's performance, we defined a figure of merit as the squared norm of the arithmetic mean of $\boldsymbol{e}_{\text{out}}^\dagger \tilde{\boldsymbol{e}}_{\text{out}}$ over all randomly selected Jones matrices and input polarizations, where † is the complex transpose operator. This figure of merit is the projection of the actual response $\tilde{\boldsymbol{e}}_{\text{out}}$ over the ideal response $\boldsymbol{e}_{\text{out}}$ and represents the polarization and phase conversion efficiency of a slowly varying metasurface that is composed of a diverse set of meta-atoms.[28] The figures of merit for different interlayer distances $S$ are shown in Figure 4b, which shows that the polarization and phase conversion efficiency is more than 84% for interlayer distances $S \geq 0.5\lambda_{SU-8}$, where $\lambda_{SU-8}$ is the wavelength in the interlayer medium, indicating that the effect of interlayer coupling is small when the layers are sufficiently distant. In this regime, the small residual reflection from the layers and differences between the desired and realized phase shifts ($\phi_1$ and $\phi_2$) result in an efficiency less than 100%.

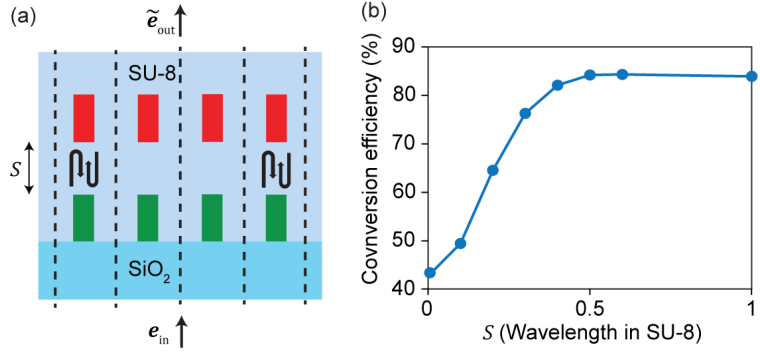

**Figure 4.** a) Coupled bilayer model. An infinite periodic array of bilayer supercells that capture effects of interlayer coupling and multiple reflections. b) Conversion efficiency of the partitioned bilayer model as a function of interlayer distance $S$.

## 4. Design Example: Bifocal Bilayer Metalens

As a demonstration of our design approach, we use the design maps in Figure 3 to design a bilayer bifocal metalens that transforms two orthogonally-polarized inputs independently: left-hand circularly polarized (LCP) light is transformed into elliptically polarized light with elevation and azimuth angles of $2\chi = 40°$ and $2\psi = 120°$ on the Poincaré sphere, while right-hand circularly polarized (RCP) light is transformed into elliptically polarized light with elevation and azimuth angles of $2\chi = -40°$ and $2\psi = 300°$. Each polarization is focused to a different focal point **(Figure 5a)**.

The diameter of the lens is 40 µm, and the distance between the layers was selected as $S = \lambda_{SU-8} = 544$ nm. The focal points are separated by a distance $d = 20$ µm in a plane at a distance $f = 160$ µm from the device aperture (Figure 5a). As mentioned above, each unit cell in the design has two residual degrees of freedom; in the design presented in Figure 5, we used these degrees of freedom to select $\theta_1 = 1.31$ rad and $\bar{\phi}_1 = 0$ rad for the first layer. The distribution patterns of the meta-atoms are shown in Figure 5b.

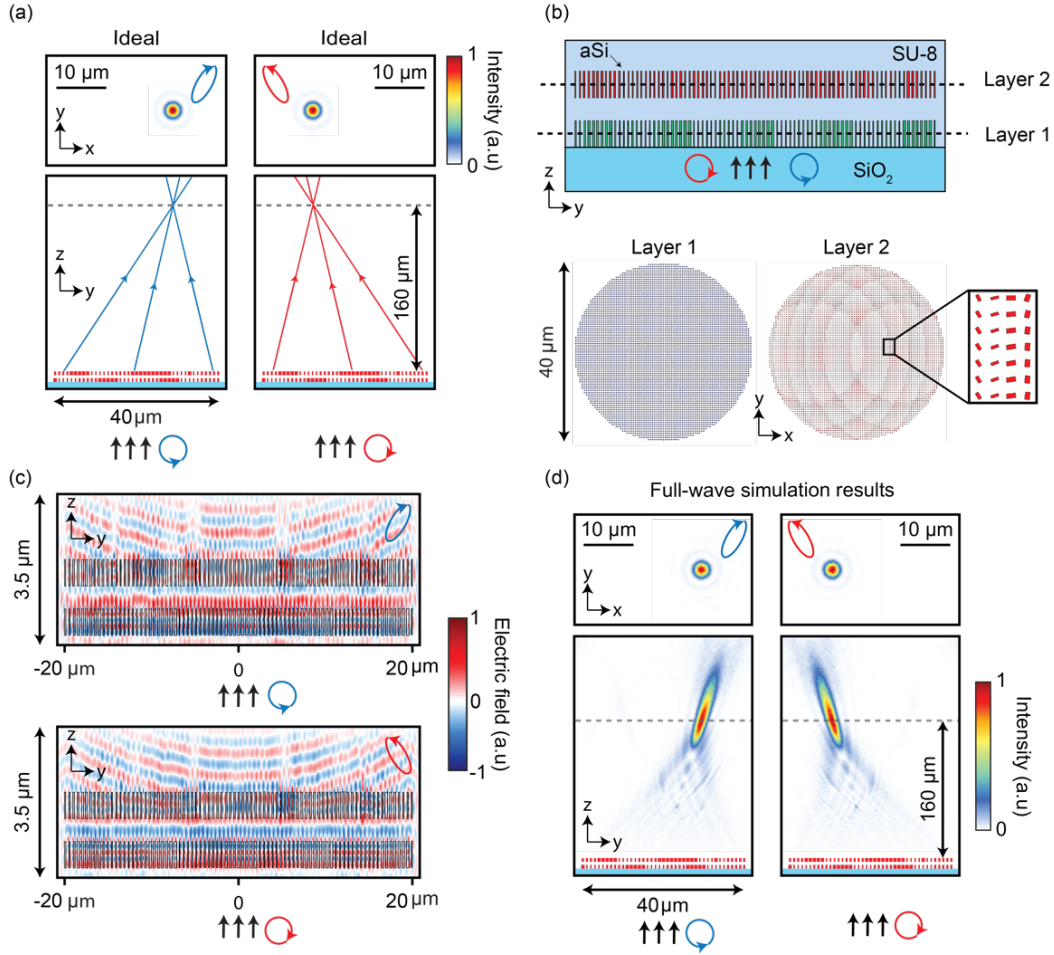

**Figure 5.** a) Schematic illustration of a bifocal metalens and its desired focal spots. b) Cross-sections and distribution of rectangular meta-atoms in each layer of a 40-μm-diameter metalens in the two planes that are shown by dashed lines. c) Simulated snapshots of the $x$ component of the electric field for the bifocal lens at each polarization state. The metalens is illuminated by normally incident circularly polarized plane waves. d) Simulated intensity distributions in the cross-sectional and focal planes for two different input circular polarizations.

To evaluate the performance of our design, we simulated the structure using an FDTD solver.[29] Figure 5c shows the simulated snapshots of the electric field in the $yz$ plane. Longitudinal cuts of the intensity in the image space for the two different orthogonal polarizations are shown in Figure 5d. The focusing efficiency of the metalens, defined as the ratio of the power with the desired polarization focused to the desired point to the input power, is found to be 70.4% and 68.9%, and the transmittance 84% and 82% for LCP and RCP incident light, respectively. The efficiency values were obtained by comparing the power passing through an 8-μm-radius aperture centered at the desired focal point to the power passing through the same aperture for an ideal metalens. In a metalens that focuses in air, the small Fresnel reflection at the SU-8/air boundary would reduce the transmission efficiency, but could be corrected using an antireflective coating. Two additional polarization multiplexing designs are presented in Figure S1 and S2.

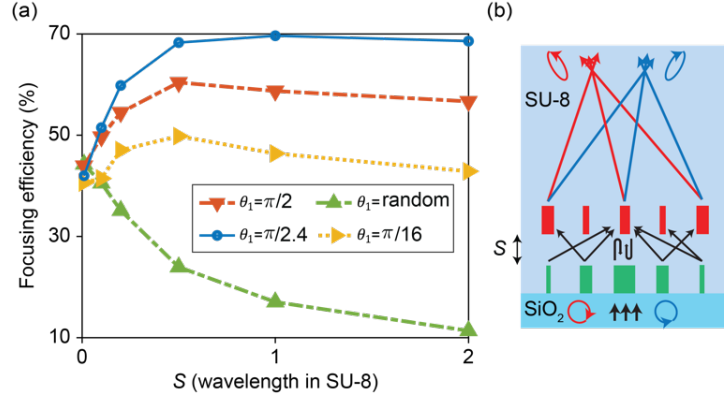

**Figure 6.** a) Simulated focusing efficiencies of four different implementations of the same metalens as a function of interlayer spacer thicknesses $S$. The eigenaxis angle of the first layer $\theta_1$ is selected as one of the free parameters and the average phase of the first layer $\bar{\phi}_1$ is assumed to be zero as the second free parameter. b) Schematic illustration of a bilayer bifocal lens showing multiple scattering effects between non-uniform neighbor meta-atoms in different layers.

There are multiple solutions for each LB decomposition due to the two residual degrees of freedom discussed in section 2, and the way solutions are chosen impacts the efficiency of a device. **Figure 6a** shows the focusing efficiency as a function of interlayer distance $S$ for designs that fix the eigenaxis angle and average phase of the first layer $\bar{\phi}_1 = 0$. For $S$ values smaller than half of the wavelength in the spacer layer, the near-field coupling between the layers reduces the efficiency. The reduction in the device efficiency for large $S$ values and the variance of efficiency among designs can be attributed to interactions between dissimilar neighboring nano-posts (see Figure 6b) that are not captured in the analysis of the periodic structures presented in Figure 4. Field cross-sections like those presented in Fig. 5c showing the bifocal lens design with different values of $S$ are shown in Fig. S3.

Solutions that minimize the local variations of unit cells would better satisfy the assumptions of the conventional design model and produce better-performing devices. The device with the highest efficiency, which is the same as the device presented in Figure 5, has a uniform first layer, while the design obtained by randomly selecting $\theta_1$ for different nano-post in the first layer (Figure 6a, green curve) has the largest local variations among the meta-atoms. Designs with uniform first layers are possible only when the incident and output light polarizations do not vary spatially, as in the metalens design we considered here. In such cases, free parameters $\theta_1$ and $\phi_1$ can be selected such that for all the meta-atoms, the first layer transforms $\boldsymbol{e}_{\text{in}}$ to an intermediate state that is the mirror image of $e_{out}$ with respect to the equatorial plane, and thus the first layer becomes uniform. The second layer is then designed to transform the mirror image of $e_{out}$ to $e_{out}$ while imparting two different arbitrary wavefronts to $e_{out}$ and its orthogonal counterpart (i.e., its antipodal state on the Poincaré sphere). It has been previously shown that the latter operation can be realized by a single layer LB element.[5]

## 5. Conclusion

Bilayers of LB meta-atoms can implement arbitrary unitary transformations that are not achievable using single layers of such meta-atoms. The geometrical analysis using the Poincaré sphere offers an intuitive understanding of the polarization transformation by the bilayer LB meta-atoms and illustrates the degrees of freedom present in LB decomposition. Decomposition of an SU(2) transformation into products of a fixed subset of transformations has been studied before in the contexts of quantum logic gates[30] and polarization gadgets formed of quarter and half waveplates.[31] However, constraints in those disciplines differ from those of the present work: here we are not limited to elements with fixed retardance, but we do need control over the average phase to implement high resolution spatially varying phase profiles.

The studies of efficiencies using periodic bilayer structures and the bifocal metalens example show that efficient metasurface devices can be realized using the proposed partitioning approach. For achieving high efficiency, a metasurface platform such as the high index contrast nano-posts should be used for the LB element implementation, and the gap between the layers should be selected to be close to the half the wavelength in the spacer layer. In addition, the choice of free parameters for LB decomposition influences the efficiency of the device, and care must be taken to reduce local variations of meta-atom geometries. The proposed partitioning technique and the bilayer platform provide a systematic and practical design method that can be used for the realization of efficient metasurfaces implementing the most general form of lossless phase and polarization transformations.

**Acknowledgements**
This work was funded by the Samsung GRO Program.

# Supporting Information

Here we describe how to decompose an arbitrary unitary polarization and phase transformation into two LB transformations. A unitary transformation can be represented by a unitary Jones matrix $\mathbf{T}^U$ that can be expressed in the $xy$ basis as

$$\mathbf{T}^U = e^{i\bar{\phi}} \begin{bmatrix} a & b \\ -b^* & a^* \end{bmatrix}, \tag{A1}$$

where $\bar{\phi}$ is the average phase delay, the complex-valued elements $a$ and $b$ satisfy $|a|^2 + |b|^2 = 1$, and * represents the complex conjugate. To determine the LB decomposition, we begin by making the Ansatz that any unitary matrix is decomposable into two unitary and symmetric matrices

$$\mathbf{T}^U = \mathbf{T}_2^{US} \mathbf{T}_1^{US}. \tag{A2}$$

Here $\mathbf{T}_1^{US}$ and $\mathbf{T}_2^{US}$ are $2 \times 2$ unitary and symmetric matrices that are decomposable in terms of their eigenvalues and linearly polarized eigenvectors as[5]

$$\mathbf{T}_m^{US} = e^{i\bar{\phi}_m}[\mathbf{e}_1^m \ \mathbf{e}_2^m] \begin{bmatrix} e^{-i\frac{\Delta\phi_m}{2}} & 0 \\ 0 & e^{i\frac{\Delta\phi_m}{2}} \end{bmatrix} [\mathbf{e}_1^m \ \mathbf{e}_2^m]^\dagger, \tag{A3}$$

$$\mathbf{e}_1^m = \begin{bmatrix} \cos\theta_m \\ \sin\theta_m \end{bmatrix}, \quad \mathbf{e}_2^m = \begin{bmatrix} -\sin\theta_m \\ \cos\theta_m \end{bmatrix}, \tag{A4}$$

for $m = 1, 2$. Here $\mathbf{e}_1^m$ and $\mathbf{e}_2^m$ are the linear eigenpolarizations, $2\theta_m$ is the azimuth angle of the eigenaxis measured on the Poincaré sphere (see Figure 1a), $\bar{\phi}_m$ and $\Delta\phi_m$ are the average phase and retardance of $\mathbf{T}_m^{US}$, and † represents the conjugate transpose operation.

We now illustrate some conditions on the solutions to the LB decomposition. We can rearrange Equation A2 to solve for $\mathbf{T}_2^{US}$ and $\mathbf{T}_1^{US}$ as

$$\mathbf{T}_2^{US} = \mathbf{T}^U [\mathbf{T}_1^{US}]^{-1}, \tag{A5}$$

$$\mathbf{T}_1^{US} = [\mathbf{T}_2^{US}]^{-1} \mathbf{T}^U. \tag{A6}$$

The matrix expressions on the right-hand sides of Equation A5 and A6 must be unitary and symmetric. By expressing $\mathbf{T}_1^{US}$ and $\mathbf{T}_2^{US}$ in their eigen decomposed forms shown in A3 and A4, expanding the matrix expression and enforcing the symmetry condition on the resulting right-hand side matrices, we obtain

$$\tan\frac{\Delta\phi_m}{2} = (-1)^m \frac{\mathrm{Re}(b)}{\mathrm{Im}(b)\cos 2\theta_m - \mathrm{Im}(a)\sin 2\theta_m}, \quad m = 1,2 \tag{A7}$$

where $\mathrm{Re}(\cdot)$ and $\mathrm{Im}(\cdot)$ indicate the real and imaginary parts of the argument, respectively. Similarly, by applying the unitary conditions to the Equation 4a and 4b, we obtain

$$[\text{Im}(a)^2 + \text{Re}(b)^2] \tan 2\theta_1 \tan 2\theta_2 + \text{Re}(ab) \tan 2\theta_2 - \text{Re}(ab^*) \tan 2\theta_1 + |b|^2 = 0.$$

(A8)

Finally, equating the determinants of the two matrices on both sides of Equation A2 we find

$$\bar{\phi}_1 + \bar{\phi}_2 = \bar{\phi} + q\pi,$$ (A9)

where $q$ is an integer. Equation A9 indicates that the average phase shift $\bar{\phi}$ for the cascaded structure is the sum of the average phase shifts for the two constitutive layers. Note that $q$ does not represent a degree of freedom in the design because $\bar{\phi}$ and $\bar{\phi} + q\pi$ represent the same average phase.

Equation A7, A8, and A9 are the four design equations that the six design parameters (two retardances $\Delta\phi_m$, two average phase delays $\bar{\phi}_m$ and two rotation angles $\theta_m$ for $m = 1,2$) should satisfy; therefore, there are two degrees of freedom in the design. The four design parameters that appear in the graphical approach using the Poincaré sphere (i.e., $\Delta\phi_m$ and $\theta_m$ for $m = 1,2$) must satisfy the three equation A7 and A8; thus, only one of them can be selected as a degree of freedom.

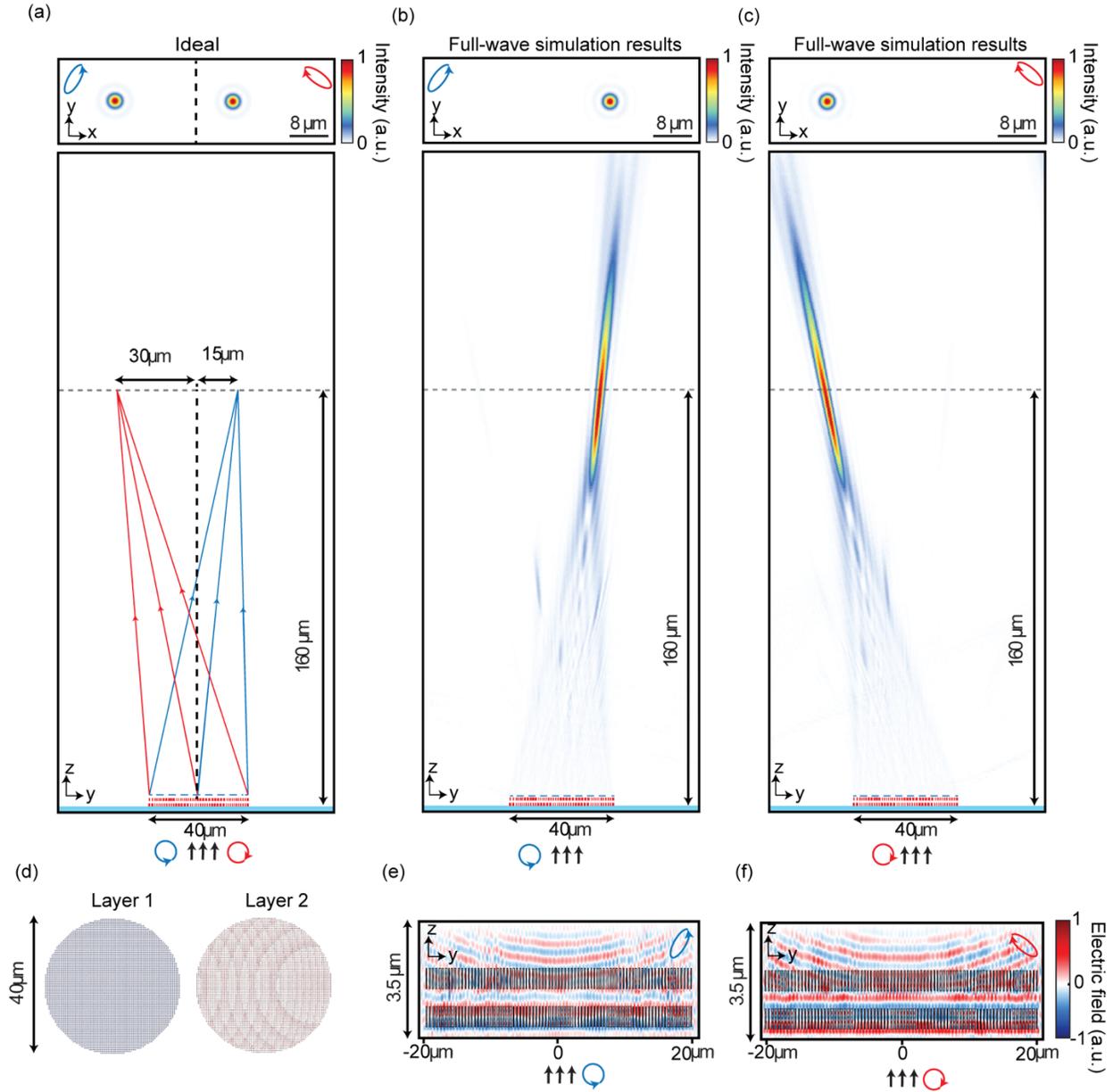

**Figure S1.** A metalens that converts LCP to $(2\chi, 2\psi) = (40°, 120°)$, focusing it to $(x, y, z) = (0, 15\ \mu m, 160\ \mu m)$; and converts RCP to $(2\chi, 2\psi) = (-40°, 300°)$, focusing it to $(x, y, z) = (0, -30\ \mu m, 160\ \mu m)$. Residual degrees of freedom are fixed by $\theta_1 =$ rad and $\bar{\phi}_1 = -$rad. The focusing efficiencies are 65.6% and 62.6% and transmittances are 83.6% and 83.5% for LCP and RCP inputs, respectively. a) Schematic illustration. b) Simulated intensity distributions in the cross-sectional and focal planes for LCP and c) RCP light. d) Distribution of meta-atoms. e) Simulated snapshots of the $x$-component of the electric field for LCP and f) RCP light.

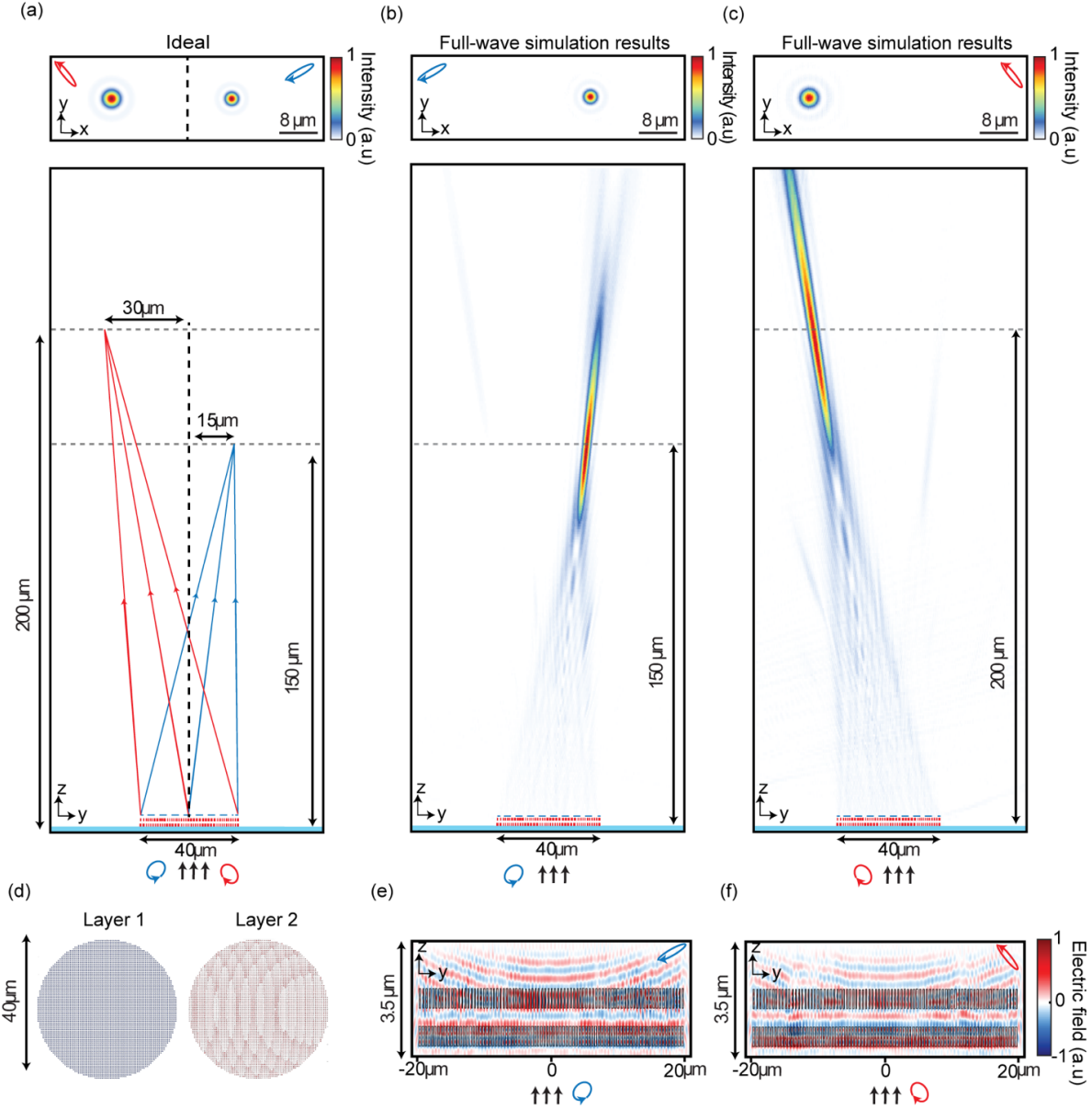

**Figure S2.** A metalens that converts $(2\chi, 2\psi) = (80°, 40°)$ to $(2\chi, 2\psi) = (-10°, 80°)$, focusing it to $(x, y, z) = (0, 15\ \mu m, 150\ \mu m)$; and converts $(2\chi, 2\psi) = (-80°, 220°)$ to $(2\chi, 2\psi) = (10°, 260°)$, focusing it to $(x, y, z) = (0, -30\ \mu m, 200\ \mu m)$. Residual degrees of freedom are fixed by $\theta_1 = $ rad and $\bar{\phi}_1 = 0$ rad. The focusing efficiencies are 63.2% and 61.4% and transmittances 80.7% and 80.4% for LCP and RCP inputs, respectively. a) Schematic illustration. b) Simulated intensity distributions in the cross-sectional and focal planes for the first and c) second polarization. d) Distribution of meta-atoms. e) Simulated snapshots of the $x$-component of the electric field for the first and f) second polarization.

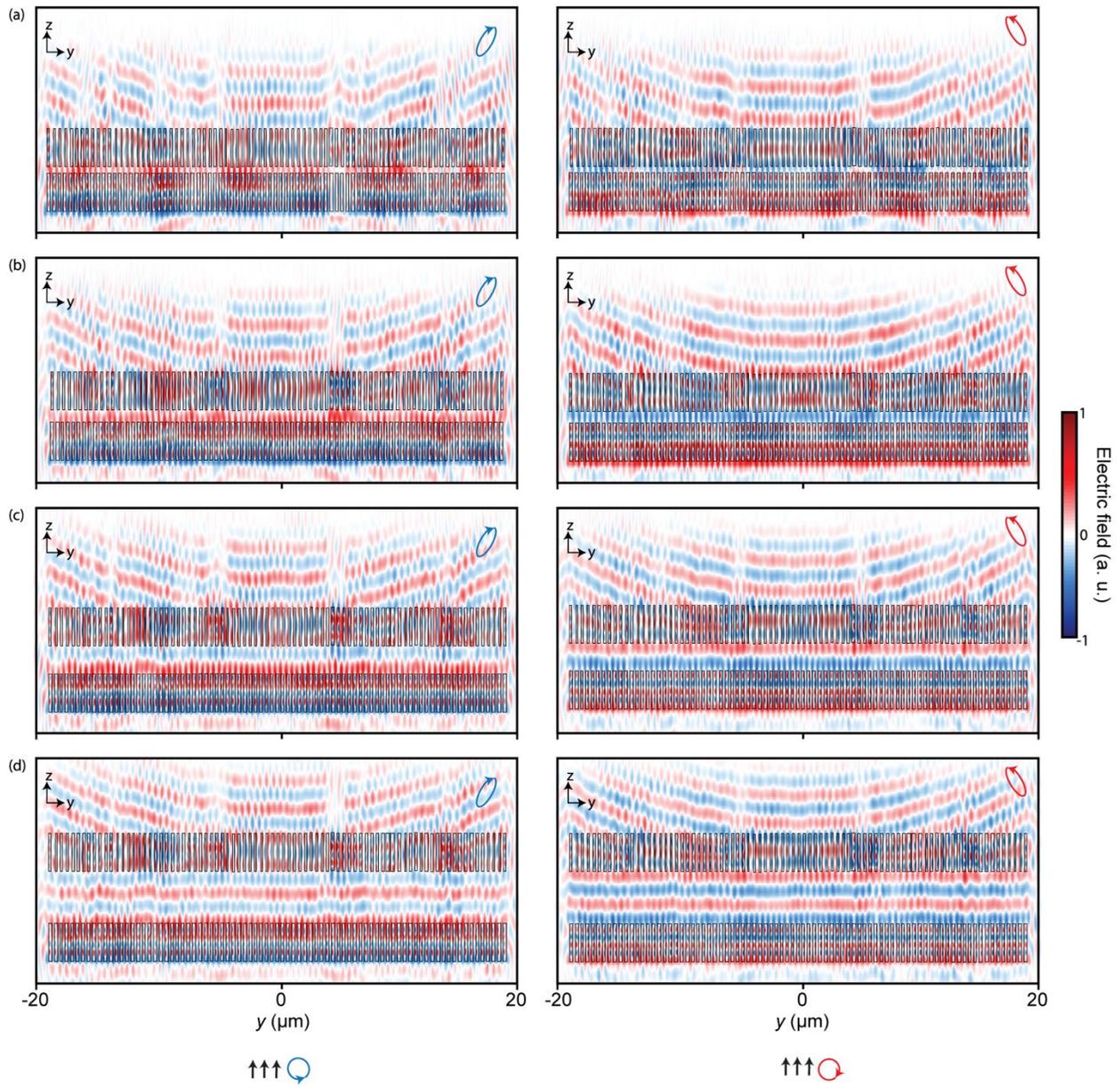

**Figure S3.** Field cross-sections for the bifocal metalens design presented in Figure 5 at varying interlayer distances. Panels on the left correspond to LCP input, and panels on the right RCP input. a) $S = 0.2\lambda_{SU-8}$. b) $S = 0.5\lambda_{SU-8}$. c) $S = 1.0\lambda_{SU-8}$. d) $S = 2.0\lambda_{SU-8}$.